\documentclass[aps,prd,twocolumn,superscriptaddress,preprintnumbers,floatfix,nofootinbib,notitlepage,showkeys,showpacs]{revtex4-1}

\usepackage[utf8]{inputenc}

\usepackage{graphicx}
\usepackage{hyperref}
\usepackage{latexsym}
\usepackage{amsmath}
\usepackage{amssymb}
\usepackage{bbm}
\usepackage{ulem}
\usepackage{pdfsync}
\usepackage{epsfig}
\usepackage{epstopdf}
\usepackage{subfigure}
\usepackage{xcolor}
\usepackage{comment}
\usepackage{slashed}
\usepackage{multirow}
\usepackage{xfrac}

\newcommand{\avg}[1]{\langle #1 \rangle}

\begin{document}

\title{Evolution of SMBHs in light of PTA measurements: implications for growth by mergers and accretion}

\author{Gabriela Sato-Polito}
\email{gsatopolito@ias.edu}
\affiliation{School of Natural Sciences, Institute for Advanced Study, Princeton, NJ 08540, United States}

\author{Matias Zaldarriaga}
\affiliation{School of Natural Sciences, Institute for Advanced Study, Princeton, NJ 08540, United States}

\author{Eliot Quataert}
\affiliation{Department of Astrophysical Sciences, Princeton University, Princeton, NJ 08544, USA}

\begin{abstract}
We study the growth of supermassive black holes accounting for both accretion and mergers. The former is informed by observations of the quasar luminosity function (QLF) and the latter by the gravitational wave-background (GWB) recently detected by PTAs, while estimates of the present-day black hole mass function provide a boundary condition. The GWB is dominated by the most massive black holes ($\gtrsim10^{9}M_{\odot}$). We show that their evolution can be simplified into a two-step process: mergers dominate at $z\leq1$, while accretion peaks at $1.4\leq z\leq2$. The large amplitude of the observed GWB suggests a significant number of mergers. We show that this generically implies a higher average Eddington ratio for quasars relative to a scenario in which mergers are negligible. In the absence of mergers, matching local estimates of BH abundance to the QLF implies a radiative efficiency $\epsilon_r=0.12$ and Eddington ratio $\lambda=0.2$. With mergers, a progenitor of mass $M_i$ is boosted to a final total mass $M_f$ and there is a direct relation between the mass gained in mergers and the average Eddington ratio of the quasar population, given by $M_f/M_i=(1+\bar{q})^N \sim \lambda/0.2$, where $\bar{q}$ is the average mass ratio and $N$ is the average number of mergers. There is thus a tension between the observed GWB, quasar properties, and the BH mass function: estimates of the mass function consistent with Eddington ratios inferred in quasars and $\epsilon_r\sim0.1$ underpredict the GWB; multiple/equal mass mergers can boost the GWB, but lead to a high Eddington ratio. If the local mass function is on the high end of current estimates, the GWB is more readily explained, but requires low efficiencies $\epsilon_r\sim10^{-2}$ not expected in standard luminous accretion models. The significant merger rate implied by the GWB also strongly suggests that the most massive BHs in the local universe have significant spin due to the orbital angular momentum from mergers, perhaps $a\sim0.5$.
\end{abstract}

\maketitle

\section{Introduction}
Supermassive black holes (SMBHs) appear to reside at the centers of all massive galaxies. Their masses can be inferred from gas or stellar dynamics methods in nearby galaxies and are found to be tightly correlated with the host properties, suggesting their co-evolution \cite{1998Natur.395A..14R, kormendy_ho}. Luminous quasars are believed to be powered by accretion onto a central supermassive black hole \cite{salpeter_1964}, which is supported by the observed evolution of AGNs and the ubiquity of remnant SMBHs in local galaxies. Understanding the formation and evolution of SMBHs, as well as its interactions with the host galaxy remains a critical question in astrophysics.

There are different approaches to theoretically model the growth of SMBHs. Analytical models rely on a continuity equation for black holes \cite{cavaliere_1971, small_blandford_1992, yu_tremaine_2002, marconi_2004, merloni_2008, SWM_2009, tucci_volonteri_2017, SWM_2013} and use the luminosity function of quasars to track the evolution of the SMBH mass function due to accretion. The simplest implementation of this model requires two free parameters: the radiative efficiency $\epsilon_r$, which relates the mass accretion rate to the luminosity, and the Eddington ratio $\lambda = L/L_{\rm edd}$, relating the black hole mass to the quasar luminosity. Beyond the simplest scenario, more involved parameter distributions, the inclusion of a merger contribution, as well as mass and redshift dependencies for the model parameters have been explored. The black hole mass function at any epoch can then be predicted by using the continuity equation to evolve it forward or backward in time. Other approaches to study the evolution and growth of SMBHs are through simulations or semi-analytic models. There are currently a plethora of cosmological-scale hydrodynamic simulations developed to self-consistently model the evolution of dark matter, galaxies, SMBHs, stars, and the interstellar medium \cite{Hirschmann:2013qfl, Springel:2017tpz, Dave:2019yyq, 2022MNRAS.513..670N}, as well as empirical models with the same goal \cite{2023MNRAS.518.2123Z}. Semi-analytic models also jointly evolve dark matter halos, galaxies, and black holes, while assuming analytic descriptions of various baryonic processes\cite{2000MNRAS.311..576K, 2003ApJ...582..559V, 2006MNRAS.370..645B, 2011MNRAS.411.1467K, 2012MNRAS.426..237H}.

The detection of nano-hertz gravitational waves by pulsar timing arrays (PTAs) \cite{NANOGrav_stoc, EPTA_stoc, PPTA_stoc, CPTA_stoc} may be offering a new observational handle on the evolution of the SMBH mass function. Following the merger of the host galaxies, SMBHs are also expected to merge, adding another contribution to the evolution of the mass function over time; since lower mass objects combine to form one of higher mass, mergers will typically reduce the number of objects below a critical mass and increase the abundance above it over time. 

In this work, we explore the evolution of the SMBH mass function including contributions from both mergers, informed by the recent PTA detections, and accretion, informed by observations of the quasar luminosity function \cite{Hopkins_2007, shen_2020}. Instead of assuming an initial condition, evolving the mass function forward in time and comparing the result at $z=0$ with local estimates of the mass function, we start with the latter as a boundary condition, since it is a relatively well measured quantity that must be matched. For a fixed mass, we show that the evolution of the most massive SMBHs ($\gtrsim 10^9 M_{\odot}$) can be simplified into a two-step process: mergers dominate the evolution for $z\lesssim 1$, while accretion peaks at $z\sim 2$. Since SMBHs can only merge later than their host halos, the halo merger rate is the highest redshift estimate of the SMBH merger. We show that for any relevant halo mass, the merger rate is dominated by $z\lesssim 1$ based on the extended Press-Schechter merger rate.

A stochastic gravitational-wave background (SGWB) produced by SMBHs suggests that they merged at least once and possibly multiple times. The simplification of the SMBH evolution at high masses into an early accretion stage followed by a later merger stage implies that the relevant mergers for the SGWB occurred after growth by accretion was largely completed. Starting from the present-day mass function, the observation of the SGWB therefore implies that black holes were less massive in the past (i.e. the abundance of low-mass black holes was larger and of high-mass black holes was smaller). Since the quasar luminosity function (QLF) is determined observationally, this necessarily implies a larger inferred Eddington ratio compared to the scenario in which mergers are insignificant. We show that the shift in the typical mass due to mergers can be approximately parametrized by the ratio of the final mass to the mass of the most massive progenitor $M_f/M_i=(1+\bar{q})^N$, where $\bar{q}$ is the average mass ratio and $N$ is the average number of mergers, and that the inferred Eddington ratio scales directly with this ratio. Using the QLF measured by Ref.~\cite{shen_2020}, we find that, in the absence of mergers, $\lambda \sim 0.2$. With mergers, we therefore have that $\lambda/0.2 \sim (1+\bar{q})^N$.

As discussed in Refs.~\cite{Sato-Polito:2023gym, Sato-Polito:2024lew}, estimates of the present-day SMBH mass function typically underpredict the amplitude of the SGWB. The amplitude can be slightly raised by requiring that black holes merge multiple times. We show, however, that there is a trade-off between the number of mergers and the average Eddington ratio that is required, which complicates this scenario, with multiple mergers quickly leading to average Eddington ratios larger than 1. If the SGWB is underpredicted due to an underestimate of the local mass function, this implies that the radiative efficiency must be $\epsilon_r\sim 10^{-2}$.

This paper is organized as follows: Secs.\ref{sec:accretion} and \ref{sec:merger} detail the evolution of the black hole mass function due to accretion and mergers, respectively. Both contributions are combined in Sec.~\ref{sec:results}, where the main results of this work are presented. In Sec.~\ref{sec:two-stage}, we point out that merger and accretion dominate at two distinct epochs for the most massive SMBHs. With this simplified description of their evolution, we show in Sec.~\ref{sec:acc+mergers} the relation between merger and accretion parameters imposed by the continuity equation and the implications of the recent PTA detection of a SGWB for SMBH evolution. In Sec.~\ref{sec:Ledd_M_z} allow the Eddington ratio to have more complicated dependencies on BH mass, redshift, or a distribution of Eddington ratios for a fixed quasar luminosity. We then conclude in Sec.~\ref{sec:conclusion}.

\section{Accretion}\label{sec:accretion}

\begin{figure}
    \centering
    \includegraphics[width=\linewidth]{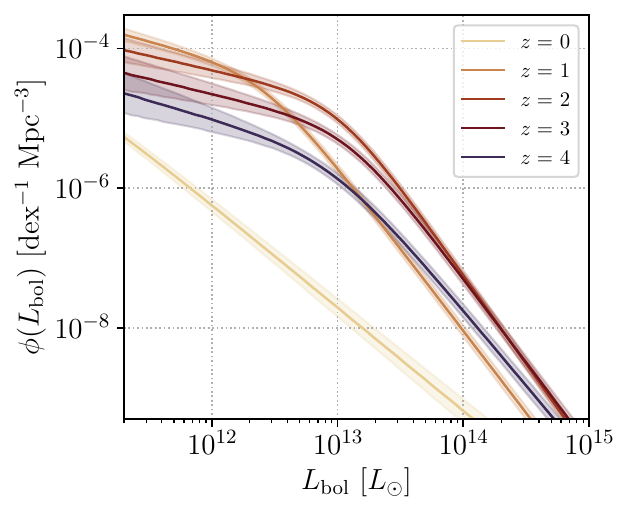}
    \caption{Quasar luminosity function for various redshifts reproduced from the fit presented in Ref.~\cite{shen_2020}.}
    \label{fig:BHMF_QLF}
\end{figure}

The evolution of the supermassive black hole mass function is commonly modelled through the continuity equation. Let $\psi(M,t) dM$ be the comoving number density of SMBHs with masses between $M$ and $M+dM$ at a time $t$, then
\begin{equation}
    \frac{\partial \psi}{\partial t}(M,t) = \Delta(M,t) + \Gamma(M,t),
    \label{eq:cont_eq}
\end{equation}
where $\Delta$ is the accretion contribution
\begin{equation}
    \Delta(M,t)= -\frac{\partial}{\partial M}\left[\avg{\dot{M}(M,t)} \psi (M,t)\right]
    \label{eq:delta}
\end{equation}
and $\avg{\dot{M}}$ is the average accretion rate over the entire black hole population (over active and inactive AGNs), and $\Gamma(M,t)$ is the change due to mergers, which we will set $=0$ for the moment, but will return to in Sec.~\ref{sec:merger}. We will assume that AGNs accrete during an active phase with a fraction $\lambda$ of the Eddington luminosity $L=\lambda L_{\rm edd}$ and convert mass into energy with an efficiency $\epsilon_r$. That is,
\begin{equation}
    L=\frac{\lambda}{t_E} Mc^2 = \epsilon_r \dot{M}_{\rm acc} c^2, 
    \label{eq:L_Macc}
\end{equation}
where $t_E=\sigma_T c/4\pi G m_p = 4.5\times 10^8$yr is the Eddington time, and $\epsilon_r \dot{M}_{\rm acc}$ corresponds to the fraction of the mass that is radiated as it falls into the black hole. The mass that is added to the black hole $\dot{M}$ is therefore $\dot{M} = (1-\epsilon_r) \dot{M}_{\rm acc}$. If we define the bolometric luminosity function of quasars $\phi(L,t) d\log_{10} L$ as the number density of quasars per log luminosity, and we suppose that a fraction $\delta(M,t)$ of black holes is active at a time $t$, then $\phi(L,t)$ is related to the black hole mass function via
\begin{equation}
    \phi(L,t) d\log_{10} L = \delta(M,t) \psi(M,t) dM.
    \label{eq:LF_MF}
\end{equation}
The quantity $\delta$ is the BH duty cycle, where $\avg{\dot{M}(M,t)} = \delta(M,t) \dot{M}(M,t)$. Combining Eqs.~\ref{eq:L_Macc} and \ref{eq:LF_MF}, we find
\begin{equation}
    \avg{\dot{M}(M,t)} \psi(M,t) = \frac{(1-\epsilon_r)}{\epsilon_r c^2 \ln(10)}\phi(L,t) \frac{dL}{dM}.
\end{equation}
we can therefore rewrite the second term on the left-hand side of Eq.~\ref{eq:cont_eq} in terms of the quasar luminosity function
\begin{equation}
    \frac{\partial \psi}{\partial t}(M,t) + \frac{(1-\epsilon_r)}{\epsilon_r} \frac{\lambda^2 c^2}{t^2_{\rm E} \ln(10)} \left. \frac{\partial \phi}{\partial L}\right|_{L=\lambda Mc^2/t_{\rm E}} = 0,
    \label{eq:cont_eq_LF}
\end{equation}
Multiplying by the mass and integrating the equation above over time shows that the black hole mass density is directly related to the integral of the quasar luminosity function
\begin{align}
    \rho_{\rm BH} &= \int dM \psi(M) M \label{eq:rhoBH} \\ &=\frac{1-\epsilon_r}{\epsilon_r c^2} \int_0^{\infty} dz \frac{dt}{dz} \int d\log_{10} L \phi(L, z) L,
\end{align}
which is the So\l{}tan argument \cite{soltan1982}.

Note that the radiative efficiency only changes an overall amplitude of $\rho_{\rm BH}$, while the Eddington ratio changes both the amplitude and the shape of the final SMBH mass function, while keeping the total mass density fixed. As a toy example consider a scenario in which the redshift evolution of the parameters is negligible, then the quasar luminosity function (QLF) is always given by a double power-law, with a break at a luminosity $L_*$. The corresponding break in the SMBH mass function will be $M_*=L_* t_E/\lambda c^2$ and the power-laws of $dn/d\log M$ will be the same as $\phi(L)$. Therefore, reducing (increasing) $\lambda$ increases (reduces) the characteristic mass of the turn-over in the SMBH mass function. In this scenario, black holes of all masses grow at the same time. The measured quasar luminosity function, however, has a break that evolves over time, mostly decreasing with redshift. This implies that more massive black holes assemble their mass earlier on. 

SMBH growth from accretion can thus be modelled by directly relating it to the observed QLF. We use the bolometric QLF presented in Ref.~\cite{shen_2020}, which is based on a compilation of various multi-wavelength observations that span IR, optical/UV, soft, and hard X-rays, and represents an update to Ref.~\cite{Hopkins_2007}. The QLF is fit by a double power-law
\begin{equation}
    \phi(L)\equiv \frac{dn}{d\log_{10} L} = \frac{\phi_*}{(L/L_*)^{\gamma_1} + (L/L_*)^{\gamma_2}}.
\end{equation}
All results shown in this work correspond to ``global fit A" in Ref.~\cite{shen_2020}, in which the model parameters $\phi_*$, $L_*$, $\gamma_1$, and $\gamma_2$ have a parametrized redshift dependence and are simultaneously fit across all redshifts.

Eq.~\ref{eq:cont_eq_LF} can be integrated over time, which results in
\begin{equation}
\begin{split}
    \psi_0(M) -\psi(M,z) =& \frac{(1-\epsilon_r)}{\epsilon_r} \frac{\lambda^2 c^2}{t^2_{\rm E} \ln(10)} \times \\ &\int_0^z dz' \frac{dt}{dz'} \left. \frac{\partial \phi}{\partial L}\right|_{L=\lambda Mc^2/t_{\rm E}},
\end{split}
\end{equation}
where $\psi_0$ is the mass function today. For a sufficiently high redshift $z$, $\psi(M,z)$ is negligible compared to $\psi_0$, i.e., in the absence of mergers, the mass function today must equal the accreted mass. We model the present-day mass function as in Refs.~\cite{Sato-Polito:2023gym, Sato-Polito:2024lew}, which we very briefly summarize here.

The present-day SMBH mass function can be estimated from scaling relations between black hole mass and properties of the host galaxy (say, property $X$) and galaxy catalogs, which give $dn/dX$. The scaling relation is expressed as
\begin{equation}
    \log_{10}M = a_{\bullet} + b_{\bullet}\log_{10} X.
    \label{eq:M-X}
\end{equation}
Using the galaxy velocity dispersion as our fiducial proxy for SMBH mass, we parametrize the velocity dispersion function (VDF) as
\begin{equation}
    \frac{dn}{d\sigma} d\sigma = \psi_{*} \left(\frac{\sigma}{\sigma_*}\right)^{\alpha} \frac{e^{-(\sigma/\sigma_*)^{\beta}}}{\Gamma(\alpha/\beta)} \beta \frac{d\sigma}{\sigma}.
    \label{eq:sigma_function}
\end{equation}
Including scatter in the relation between black hole mass and galaxy property leads to
\begin{equation}
    \psi_0(M) = \int d\sigma \frac{p(\log_{10}M|\log_{10}\sigma)}{M \log(10)} \frac{dn}{d\sigma}(\sigma),
    \label{eq:MF_scatter}
\end{equation}
where $p$ is assumed to be log-normal
\begin{equation}
\begin{split}
    p(\log_{10}&M|\log_{10}\sigma) = \frac{1}{\sqrt{2\pi} \epsilon_0} \\ &\times \exp\left[ -\frac{1}{2}\left(\frac{\log_{10}M - a_{\bullet}-b_{\bullet}\log_{10}\sigma}{\epsilon_0}\right)^2 \right].
\end{split}
\end{equation}
As in Refs.~\cite{Sato-Polito:2023gym, Sato-Polito:2024lew}, the adopt the velocity dispersion function as the mass proxy. We therefore use $X=\sigma/200$km s$^{-1}$ in Eq.~\ref{eq:M-X} and the $M-\sigma$ relation from Ref.~\cite{mcconnell_ma}, which corresponds to $a_{\bullet}= 8.32\pm 0.05$, $b_{\bullet}= 5.64\pm 0.32$, and $\epsilon_0=0.38$. For the VDF, we choose the measurement presented in Ref.~\cite{bernardi_VDF} from SDSS, fit to all galaxies with $\sigma>125$km s$^{-1}$, which corresponds to parameters: $\psi_*=(2.61 \pm 0.16)\times 10^{-2}$Mpc$^{-3}$, $\sigma_*=159.6\pm1.5$ km s$^{-1}$, $\alpha=0.41\pm 0.02$, and $\beta=2.59\pm 0.04$.

Similarly to Ref.~\cite{shankar_2009}, if we consider the QLF measured in Ref.~\cite{shen_2020}, we can find the radiative efficiency $\epsilon_r$ such that the black hole mass density matches the one inferred by local estimates of the black hole mass function. The BH mass function estimated from $M-M_{*}$ in Ref.~\cite{liepold_ma_2024} suggests $\rho_{\rm BH} = (1.8 ^{+0.8}_{-0.5}) \times 10^6$M$_{\odot}$Mpc$^{-3}$, which requires $\epsilon_r=0.03\pm 0.01$. Using the velocity dispersion yields $\rho_{\rm BH} = (4.5 \pm 0.5) \times 10^{5}$ M$_{\odot}$Mpc$^{-3}$, which requires $\epsilon_r=0.12^{+0.01}_{-0.02}$.

We can further generalize Eq.~\ref{eq:cont_eq_LF} by considering that the Eddington ratio may depend on the properties of the black hole. We consider a power-law dependence on the black hole mass, with a pivot mass $M_{\star}$ and coefficient $\alpha$, and similarly for the redshift dependence, with a pivot redshift at $z_{\star}$ and power $\beta$, where $L_{\star} = \lambda M_{\star} c^2/t_E$. That is,
\begin{equation}
    L(M,z)=L_{\star} \left(\frac{M}{M_{\star}}\right)^{\alpha} \left(\frac{1+z}{1+z_{\star}}\right)^{\beta}.
    \label{eq:L_Mz}
\end{equation}
The mass density in black holes per logarithmic mass is then given by
\begin{equation}
\begin{split}
    M\frac{dn}{d\log_{10} M} = -\frac{1-\epsilon_r}{\epsilon_r c^2} \int dz \frac{dt}{dz} \bigg\{ \alpha(\alpha -1)L \phi(L)& \\ + \alpha^2 L^2 \frac{\partial \phi}{\partial L} \bigg\}&.
    \label{eq:rhoBH_Ledd}
\end{split}
\end{equation}
However, unless otherwise specified, we assume $ \alpha=1$.

\begin{figure}
    \centering
    \includegraphics[width=0.9\linewidth]{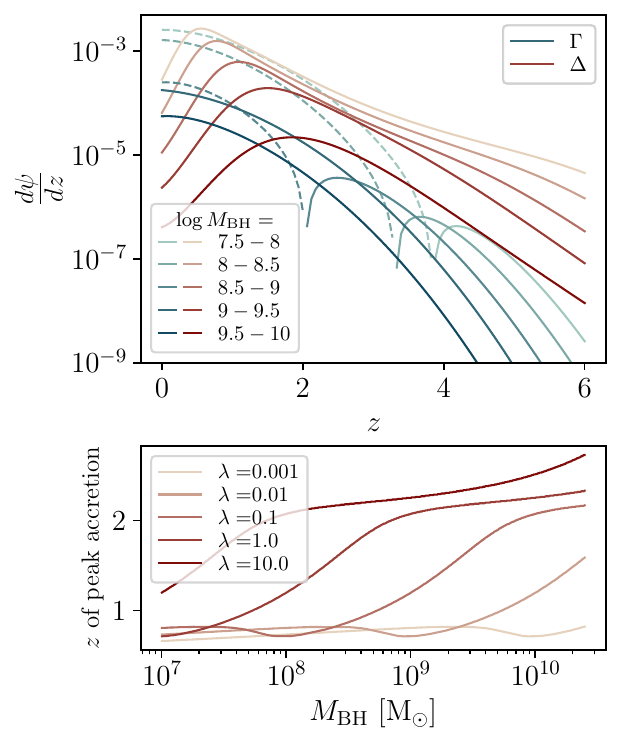}
    \caption{The top panel shows the change to the mass function per redshift due to mergers and accretion on top for different mass ranges, following Eqs.~\ref{eq:delta} and \ref{eq:smol_rate}. The dashed lines correspond to negative contributions, since mergers can reduce the abundance of black holes in certain mass ranges. The fiducial set of parameters for the relation between halo and SMBH mergers outlined in Sec.~\ref{sec:merger} were adopted, while for the accretion term, we assumed $\epsilon_r=0.12$ and $\lambda=0.2$. The bottom panel shows the redshift at which the black hole growth from accretion peaks as a function of mass and for different values of the Eddington ratio.}
    \label{fig:dpsi_dz}
\end{figure}

\section{Mergers}\label{sec:merger}
The change in the abundance of black holes of a mass $M$ due to mergers must be given by a creation rate of two black holes of lower masses $m_1$ and $M-m_1$ merging to form a black hole of total mass $M$, and a destruction rate of black holes of a mass $M$ merging with those of any other mass, that is
\begin{widetext}
\begin{equation}
\begin{split}
    \Gamma(M,t) =& \frac{1}{2}\int_0^M dm_1 \int_0^M dm_2 \delta_D(m_1+m_2-M) \psi(m_1,t) \psi(m_2,t)  Q(m_1, m_2, t) \\ 
    &- \psi(M,t)\int_0^{\infty}dm_2 \psi(m_2,t) Q(M,m_2,t),
\end{split}
\end{equation}
\end{widetext}
where $Q$ is a kernel related to the microphysical process of merger and is equivalent to a velocity times cross-section. The equation above is known as the Smoluchowski coagulation equation and was first studied in the context of halo or black hole mergers in Refs.~\cite{benson_2004, erickcek_2006}.

Notice that the merger rate density $\frac{d^3n}{dm_1 dm_2dt}$ between black holes of masses $[m_1,m_1+dm_1]$ and $[m_2,m_2+dm_2]$ are precisely the integrands above
\begin{equation}
    \frac{d^3n}{dm_1 dm_2dt} = \psi(m_1,t)\psi(m_2,t) Q(m_1,m_2,t).
\end{equation}
The first term is equivalent to the merger rate of all black hole binaries with total mass adding up to $M_t=M$. Switching from constituent masses to total mass and mass ratio $(m_1, m_2) \rightarrow (M_t, q)$, where the Jacobian is $J=M_t/(1+q)^2$, the first term can be written as 
\begin{equation}
\frac{d^2n}{dM_t dt}= \int_{q_{\rm min}}^{1} dq \frac{d^3 n}{dM_t dq dt},
\end{equation}
where the integral is limited to $q=m_2/m_1<1$ to avoid double counting the mergers that result in a total mass $M_t$. The second term is equivalent to fixing one of the masses and integrating over all values of the second mass, and we must therefore integrate over all values of $q$ and $M_t=M(1+q)$. We can now rewrite $\Gamma$ as
\begin{equation}
\begin{split}
     \Gamma(M,t) &= \frac{d^2n}{dM_t dt} - \int dM_t \int_{q_{\rm min}}^{q_{\rm max}} dq \frac{d^3n}{dM_t dq dt} \times \\ & \delta_D \left(\frac{M_t}{1+q}-M \right)
\end{split}
    \label{eq:smol_rate}
\end{equation}

The GW energy density per logarithmic frequency is related to the quantities above by
\begin{equation}
\begin{split}
  \frac{d\rho_{\rm gw}}{d\log f} (f) &= \int dt \int dM_t \frac{d^2n}{dM_t dt} \frac{1}{(1+z)}\times \\ & \left.\frac{dE^r_{\rm gw}}{d\log f_r}\right|_{f_r=(1+z)f},  
\end{split}
\end{equation}
and the characteristic strain is $h^2_c(f) = \frac{4G}{\pi f^2 c^2} \frac{d\rho_{\rm gw}}{d\log f}$.

\begin{figure*}[t]
    \centering
    \includegraphics[width=0.95\linewidth]{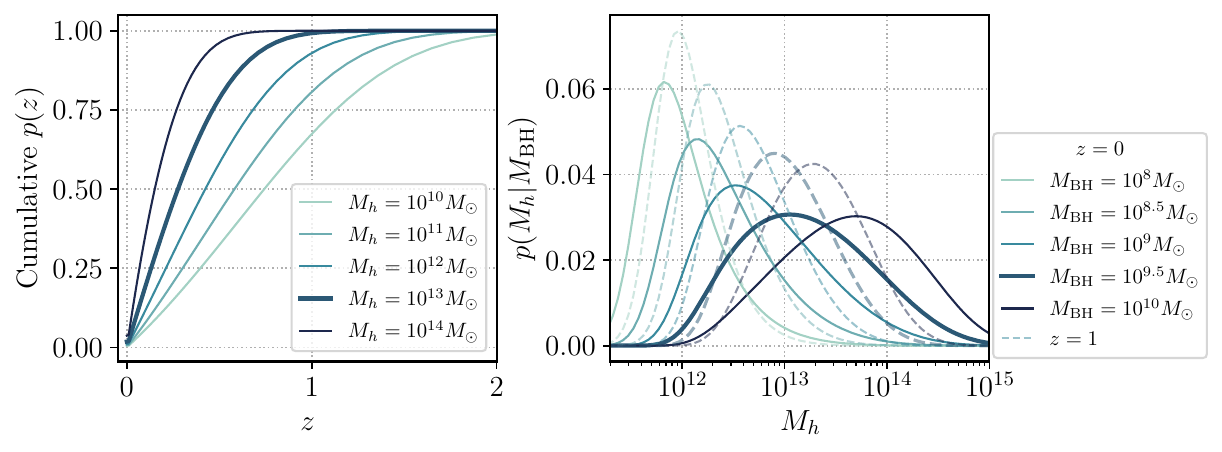}
    \caption{Redshift distribution of mergers for different halo masses and distribution of host halo masses for various SMBH masses. The left panel shows the cumulative redshift distribution computed from the excursion set formalism given in Eq.~\ref{eq:merger_rate_z}, which we detail in App.~\ref{app:EPS}. The panel on the right shows the probability of a halo of mass $M_h$ hosting a SMBH of masses spanning $M=10^8-10^{10} M_{\odot}$. That is, $p(M_h|M_{\rm BH}) = \frac{1}{\bar{N}} \int dM_* \frac{dn}{dM_h} p(M_h|M_*) p(M_*|M_{\rm BH})$. The thicker line in both plots highlights the curves that are representative of the typical SMBH that most contributes to the GW signal.}
    \label{fig:halo_merger}
\end{figure*}

For the results shown in Figs.~\ref{fig:dpsi_dz} and \ref{fig:halo_merger}, we compute the SMBH merger rate from the halo merger rate derived from the excursion set formalism outlined in App.~\ref{app:EPS}. The SMBH merger rate is given by (similar to, e.g., \cite{Ellis:2023dgf, Ellis:2023owy})
\begin{equation}
\begin{split}
    \frac{d^3 n}{dm_1 dm_2 dt} =& \int dM_{h,1} dM_{h,2}\ p(m_1|M_{h,1}, z)\times \\ & p(m_2|M_{h,2}, z) \frac{d^3 n}{dM_{h,1} dM_{h,2} dt},
\end{split}
\end{equation}
where $p(m|M_h, z)$ captures the relation between the black hole and halo mass as a function of redshift. To compute this quantity, we convert the halo mass to stellar mass as a function of redshift using the relations provided in Ref.~\cite{2020A&A...634A.135G} and a bulge stellar mass to black hole mass relation from Ref.~\cite{mcconnell_ma}. To relate the total stellar mass ($M_{*}$) to the bulge mass, we use the relation adopted in Ref.~\cite{Chen:2018znx}, based on Refs.~\cite{2014MNRAS.443..874B, Sesana:2016yky}. We further adopt a log-normal scatter in the $M_h-M_{*}$ and in the $M_b-M_{\bullet}$ with fiducial values of $\epsilon_{*} =0.2$ and $\epsilon_{\bullet} = 0.4$, i.e.
\begin{equation}
\begin{split}
    p(\log X|\log Y) =& \frac{1}{\sqrt{2\pi} \epsilon} \exp\left[ -\frac{1}{2}\left(\frac{\log X - \log \bar{X}(Y)}{\epsilon}\right)^2 \right],
\end{split}
\end{equation}
for both relations. 

We emphasize that, while the merger rate based on halos and their prescribed connection to galaxies and SMBHs is used in Figs.~\ref{fig:dpsi_dz} and \ref{fig:halo_merger}, the main results of this work, which we present in Secs.~\ref{sec:acc+mergers} and \ref{sec:Ledd_M_z}, are independent of it. Ultimately, we show that mergers dominate at lower redshifts than accretion, and therefore adopt a simpler description in which only the shift from the initial mass of the most massive progenitor to the mass of the final black hole is relevant.

\section{Evolution with mergers and accretion}\label{sec:results}
\subsection{Two-stage evolution} \label{sec:two-stage}
Fig.~\ref{fig:dpsi_dz} shows the change in the SMBH mass function due to mergers and accretion. The top panel shows that low redshifts always dominate the merger contribution, while accretion typically peaks around $z\sim 2$ for the most massive black holes. This suggests that SMBH evolution (for the most massive objects) can be simplified by dividing it into two epochs: at low redshifts ($z\lesssim 1$) mergers dominate the evolution of the mass function, while at $z\gtrsim 1$ accretion dominates.

While the results of Fig.~\ref{fig:dpsi_dz} are based on the fiducial set of choices presented above, we argue that the mergers that contribute to the SGWB can be robustly predicted to occur below $z\sim 1$. In App.~\ref{app:EPS}, we review the halo merger rate derived from extended Press-Schechter and show that the redshift distribution for a merger of total halo mass $M_{h,t}$ is given by
\begin{equation}
    p(z|M_{h,t})\propto \delta_c(z) \frac{d\delta_c}{dz} e^{-\delta_c^2/2\sigma^2(M_{h,t})},
    \label{eq:merger_rate_z}
\end{equation}
where $\delta_c(z)\approx 1.686/D(z)$ is the critical overdensity for collapse and $D(z)$ is growth factor. The cumulative redshift distribution is shown on the left panel of Fig.~\ref{fig:halo_merger} for various total masses. Note that, while the halo merger rate derived from EPS predicts a particular dependence on the mass ratio, it does not affect the redshift distribution; mergers of any mass ratio have the same redshift distribution. We also note that the number of mergers per halo of total mass $M_{h,t}$ divides out the exponential dependence on mass. The remaining factor can be rewritten as $\delta_c(z) \frac{d\delta_c}{dz} = \delta^2_c(z) f/(1+z)$, where $f\equiv \frac{d\ln D}{d\ln a}$ is the growth rate. The number of mergers per halo therefore increases with redshift, but the merger rate is suppressed for halos above the characteristic mass $\delta_c(z)=\sigma(M_c)$.

The right panel of Fig.~\ref{fig:halo_merger} shows that the typical host halo of the SMBHs that dominate the SGWB ($M_{\rm peak} \sim 3 \times 10^9 M_{\odot}$) has a mass of roughly $M_h \sim 10^{13} M_{\odot}$ and above, while the left panel shows that the merger rate for halos of mass $M_h \gtrsim 10^{13} M_{\odot}$ is completely dominated below redshift 1. Even for significantly lower halo masses, such as $M_h = 10^{10} M_{\odot}$, all mergers occur below $z=2$ and over $65\%$ below $z=1$. We therefore conclude that for any reasonable halo-SMBH relation, sources that significantly contribute to the SGWB will originate from $z\lesssim 1$. The inclusion of any time delay between halos, galaxy, and SMBHs merging will only lead to the signal being dominated by even lower redshifts.

In order to deviate from the conclusion above, Fig.~\ref{fig:halo_merger} suggests that either the peak mass that contributes to the SGWB must be significantly lower (i.e. by over two orders of magnitude) or that massive SMBHs must be hosted by significantly lower halo masses in the past. In the first scenario, producing a SGWB that agrees with PTA measurements would become even more challenging and would imply a larger discrepancy between estimates of the present-day mass function and PTAs, due to the upper limit discussion in Ref.~\cite{Sato-Polito:2023gym}. In the second scenario, the host halo mass must shift by over three orders of magnitude between $z=0-2$.

\subsection{Constraints from quasars, GWB, and present-day mass function}\label{sec:acc+mergers}

\begin{figure}
    \centering
    \includegraphics[width=0.9\linewidth]{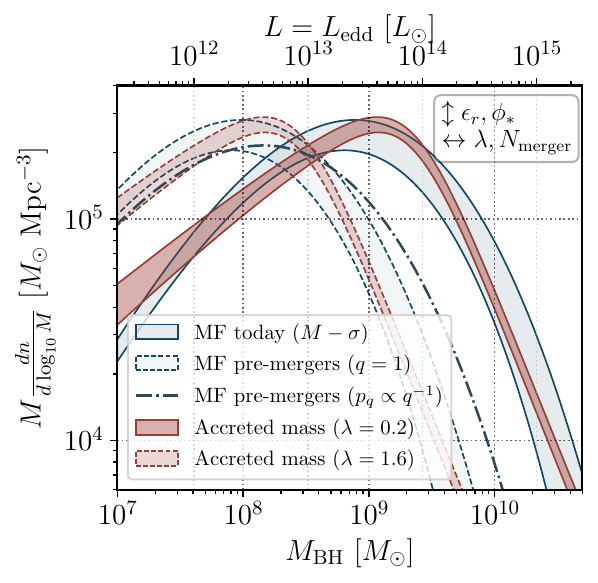}
    \caption{Mass density per logarithmic mass as a function of black hole mass. The blue curves show the mass density inferred from local estimates of the BH mass function based scaling relations, while the red are those inferred from the integrated quasar luminosity function for a fixed Eddington ratio. The solid shaded bands correspond to the mass density in the absence of mergers and the corresponding Eddington ratio required for the local and quasar-based inferences to agree, while the dashed shaded bands show the same, but assuming $N=3$ equal mass mergers. The dash-dotted line shows the result for $N=3$ mergers assuming a mass ratio distributed as $p_q(q)\propto q^{-1}$ and a minimum value of $q_{\rm min}=0.1$.}
    \label{fig:mass_density}
\end{figure}

\begin{figure}[t]
    \centering
    \includegraphics[width=0.9\linewidth]{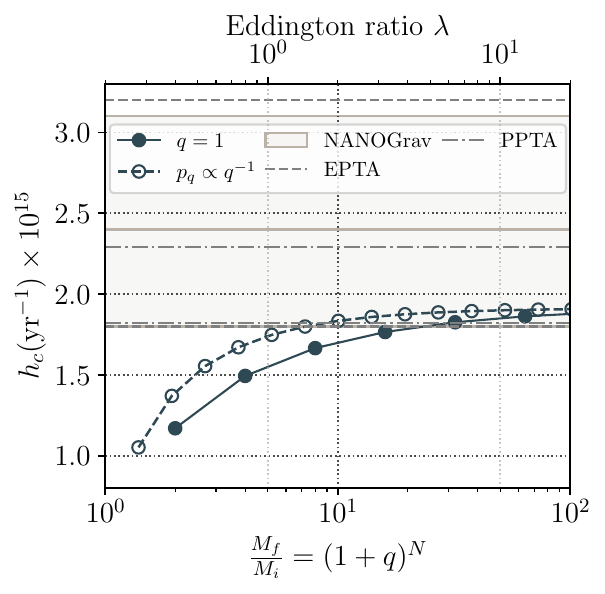}
    \caption{Characteristic strain produced by multiple merger events as a function of the ratio between the final mass and the mass of the most massive progenitor. The solid line corresponds to the equal mass case, where the filled dots show the integer number of mergers, while the dashed line and unfilled circles show the same but for a log-uniform mass ratio distribution. The shaded gray band shows the 90\% confidence interval measured by NANOGrav, the gray dashed lines for EPTA, and the dashed-dotted shows the 68\% confidence interval for PPTA. The upper x-axis shows the Eddington ratio required to match the accreted mass inferred from the quasar luminosity function and the local mass function, assuming a certain number $N$ of merger events.}
    \label{fig:hc_Ledd}
\end{figure}

The two-stage evolution discussed above therefore suggests that in order to connect the present-day mass function and the GWB signal to quasar observations we only need to predict the mass function prior to mergers, and the only requirement is that the accreted mass function inferred from the quasar luminosity function must match the pre-merger SMBH mass function. Instead of modeling the merger rate of SMBHs by connecting it to the halo merger rate as computed above, which depends on the details of the SMBH-galaxy-halo connection, we choose a simpler parametrization of the redshift and mass ratio, and compute the mass function before mergers as a function of the number of mergers.

If we suppose that all black holes merged once according to a merger rate as a function of component masses $d^2n/dm_1 dm_2$, then the mass function prior to merging is the integral of the merger rate over one of the masses
\begin{equation}
\begin{split}
\psi_1 (M) =& \int dm_1 dm_2 \frac{d^2n}{dm_1 dm_2} \delta_D(m_1-M)\\
=& \int_0^1 dq (1+q) \frac{d^2n}{dM_t dq} (M_t=(1+q)M) \\ &+ \int_0^1 dq \frac{(1+q)}{q}\frac{d^2n}{dM_t dq} \left(M_t=\frac{(1+q)}{q}M\right),
\label{eq:MF_pre-merg}
\end{split}
\end{equation}
where the the first integral in the second line corresponds to the mergers with $m_1>m_2$ and the second integral to $m_2>m_1$. We assume a merger rate described by
\begin{equation}
    \frac{dn}{dM_t dq} (M,q) = p_q(q)\psi_0(M).
\end{equation}
Eq.~\ref{eq:MF_pre-merg} follows the intuition that mergers combine a black hole of mass $M/(1+q)$ with $M q/(1+q)$ to form one of mass $M$. If the final mass function $\psi_0$ is known, we can relabel $M/(1+q) \rightarrow M$ and conclude that the abundance $\psi_1$ of black holes of mass $M$ prior to the latest merger will be the sum of the number of black holes that ended with masses $M(1+q)$ and $M(1+q)/q$ after the latest merger. In the equal mass scenario, the mass function prior to $N$ mergers is given by
\begin{equation}
    \psi_N(M) = 2^{2N} \psi_0(2^{N} M).
\end{equation}

Observations of the quasar luminosity function directly constrain the mass density accreted onto supermassive black holes. Requiring that the total accreted mass density matches the one inferred from the present-day mass function fixes the radiative efficiency, as discussed in Sec.~\ref{sec:accretion}. Hence, while the total mass density can be determined observationally, modelling the evolution of the BH mass function requires a relation between the quasar luminosity and the mass of the black hole powering it, which is given by the Eddington ratio. 

\begin{figure*}
    \centering
    \includegraphics[width=0.8\linewidth]{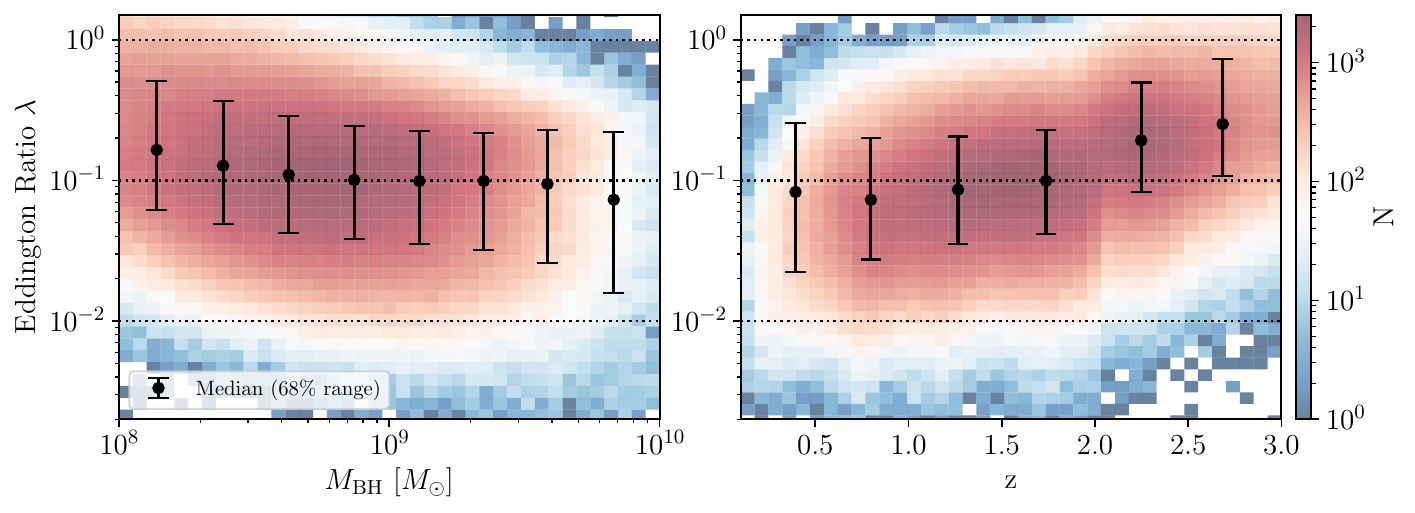}
    \caption{Derived Eddington ratios from Ref.~\cite{QSO_SDSS} in bins of BH mass (on the left) and redshift (on the right). The 2d-histogram shows the number of quasars in each bin, while the points with error bars correspond to the median and 68\% confidence intervals. Fiducial masses for $z<0.7$ are based on the H$\beta$ line, $0.7\leq z < 2$ are based on Mg II, and C IV for $z\geq 2$. The Eddington ratios are shown as a point of comparison, in order to put the values discussed in Sec.~\ref{sec:acc+mergers} in perspective.}
    \label{fig:Ledd_SDSS}
\end{figure*}

Fig.~\ref{fig:mass_density} shows the black hole mass density per logarithmic mass bin inferred from scaling relations and from the quasar luminosity function for $\alpha=1$. Since the Eddington ratio only changes the relation between quasar luminosity and BH mass, its only effect is to move the curve along the x-axis in Fig.~\ref{fig:mass_density}, while keeping the total mass density constant. Mergers have the same effect on the mass function, since they only redistribute the black hole masses, under the approximation that mass is conserved during mergers. 

In the absence of mergers, the present-day black hole mass function can be directly compared to the accreted mass inferred from the quasar luminosity function (e.g., similar to Ref.~\cite{shankar_2009}). We find a consistent total black hole mass density for $\epsilon_r=0.12$ and the shape of the mass function for $\lambda=0.2\pm 0.05$, shown in the solid bands in Fig.~\ref{fig:mass_density}. In the presence of mergers, the total mass density is conserved, but shifts the curve to the left due to the increased abundance of low mass black holes, shown in the dashed shaded bands. Since the number of massive black holes has decreased, but the mass function must match the same quasar luminosity function, this generally requires a larger Eddington ratio. In the equal mass scenario, the required Eddington ratio is related to the number of mergers $N$ via
\begin{equation}
    \frac{\lambda}{0.2} \sim 2^N.
    \label{eq:Ledd_Nmerger}
\end{equation}
The dashed curves in Fig.~\ref{fig:mass_density} correspond to $N=3$, while the dashdotted curve shows the result for a mass ratio distribution given by $p_q(q)\propto q^{-1}$ and with minimum value of $q_{\rm min}=0.1$. Similarly to the result found in the App.~A of Ref.~\cite{Sato-Polito:2023gym}, the predicted mass function for the full $q$ distribution is extremely well approximated by the average mass ratio $\bar{q}$. For the case above, $\bar{q} = 0.39$. The most massive progenitor of a BH of mass $M$ on average has a mass of $M/(1+\bar{q})^N$ and the shift of the mass density approximately follows the ratio of the final to the inital mass. Hence, Eq.~\ref{eq:Ledd_Nmerger} becomes 
\begin{equation}
\frac{\lambda}{0.2} \sim (1+\bar{q})^N
\label{eq:Ledd_Nmerger_general}
\end{equation}
The fiducial value of $N=3$ was chosen in the plot as it corresponds to the average number of mergers of a $10^{13} M_{\odot}$ halo, but we treat the relevant parameter $(1+\bar{q})^N$ essentially as a free parameter in the following discussion.


We can therefore conclude that there is a direct relation between the number of mergers and the average Eddington ratio. As shown in Ref.~\cite{Sato-Polito:2023gym}, estimates of the local SMBH mass function typically lead to an underprediction of the SGWB when compared to the measurements reported by PTAs. If the amplitude of the background is raised by requiring that black holes merge multiple times, we show in Fig.~\ref{fig:hc_Ledd} that in order to reach the 90\% lower bound of the measurement reported by NANOGrav, an Eddington ratio $\lambda>1$ is required. We highlight, however, that Ref.~\cite{Liepold:2024woa} reported a new measurement of the stellar mass function that leads to a higher amplitude. Since it is currently unclear what is the resolution of this discrepancy, we report results for both estimates.

For the sake of comparison, we show the distributions of Eddington ratio as a function of black hole mass and redshift obtained from Ref.~\cite{QSO_SDSS} in Fig.~\ref{fig:Ledd_SDSS}. The bolometric luminosity is estimated from the measured continuum luminosity at three potential rest wavelengths, depending on the redshift, while the black hole mass is estimated from recipes based on H$\beta$, Mg II, and C IV lines. We refer the reader to Ref.~\cite{QSO_SDSS} for further details and, e.g., Ref.~\cite{2013BASI...41...61S} for a discussion of the challenges associated with such estimates. Fig.~\ref{fig:Ledd_SDSS} indicates that there is no significant evolution in mass or redshift, and that the typical values for the Eddington ratio are around $\lambda\sim 0.1$. This also suggests that a small number of mergers is sufficient to produce an Eddington ratio that exceeds the typical values found by this method. If indeed multiple merger events are required to produce the observed SGWB, this may suggest an underestimate of black hole masses from line widths.

\begin{figure}
    \centering
    \includegraphics[width=0.95\linewidth]{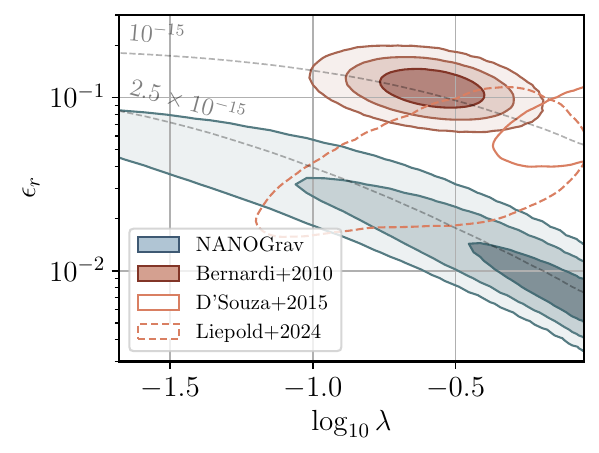}
    \caption{Constraints on $\rho_{\rm BH}-M_{\rm peak}$ derived in Ref.~\cite{Sato-Polito:2024lew}, converted into the $\epsilon_r-\lambda$ parameter space. In order to perform this conversion, we assume a value of $(1+\bar{q})^N = 1.39$, which is equivalent to all BHs having undergone a single merger event and the fiducial mass ratio distribution described in Sec.~\ref{sec:acc+mergers}. The shaded contours correspond to 1-, 2-, and 3-$\sigma$ confidence intervals, while the unshaded ones correspond to 2-$\sigma$. The values required to fit the NANOGrav 15-yr free-spectrum posterior are shown in blue, while the shaded red contour corresponds to the values inferred from the local SMBH mass function derived from the $M-\sigma$ relation. The unfilled orange contours correspond to alternative estimates of the local mass function from Refs.~\cite{DS15, liepold_ma_2024}.}
    \label{fig:eps_lamb_contour}
\end{figure}

Finally, we revisit the analysis presented in Ref.~\cite{Sato-Polito:2024lew} and translate the posteriors shown in Fig.~5 into the $\epsilon_r-\lambda$ parameter space, shown in Fig.~\ref{fig:eps_lamb_contour}. The constraints shown in Fig.~5 of Ref.~\cite{Sato-Polito:2024lew} are derived by refitting the 15-yr NANOGrav free-spectrum posterior with a model for the background given by the probability distribution of characteristic strains $p(h^2_c)$, which accounts for Poisson fluctuations in the number of sources. For the scaling relation, we adopt the fiducial model based on the $M-\sigma$ relation.

In order to convert the contours, we assume that all black holes merged once, with a value of $(1+\bar{q})^N = 1.39$, and then perform a change of variables: we use that the black hole mass density is $\rho_{\rm BH} \propto (1-\epsilon_r)/\epsilon_r$ and that $M_{\rm peak} \propto \lambda^{-1}$. The former follows directly from the Soltan argument (see Eq.~\ref{eq:rhoBH}), while the latter follows from the relation between black hole mass and luminosity. Consider the mass kernel for the characteristic strain obtained from Eq.~\ref{eq:rhoBH_Ledd}
\begin{equation}
    M^{5/3}\frac{dn}{d\log_{10} M} \propto \int dt L^{8/3} \frac{\partial \phi}{\partial L},
\end{equation}
for $\alpha=1$. The right-hand side will have a maximum at some value $L_{\rm peak}$, which will correspond to a mass $M_{\rm peak} = L_{\rm peak} t_E/\lambda c^2$. Hence the scaling of the peak mass as $M_{\rm peak} \propto \lambda^{-1}$.

For comparison, we also show in Fig.~\ref{fig:eps_lamb_contour} values of $\epsilon_r$ and $\lambda$ consistent with other approaches to model the present-day mass function based on the relation between bulge mass and black hole mass. We include the stellar mass function from Ref.~\cite{DS15} and \cite{liepold_ma_2024}, and the $M-M_b$ relation from Ref.~\cite{mcconnell_ma}. We sample the posterior of the mass function from Ref.~\cite{liepold_ma_2024} using the code provided by the authors in Appendix~B, while Ref.~\cite{DS15} does not include uncertainties on the mass function parameters and therefore the contour only includes uncertainties in the scaling relation. Note that, beyond the aforementioned assumptions regarding the mass ratio distribution and the number of mergers, the two predictions based on the $M-M_b$ relation require the translation between total stellar mass and bulge mass, which we assume here to be one-to-one. This can significantly bias the results shown in Fig.~\ref{fig:eps_lamb_contour} towards lower radiative efficiencies and higher Eddington ratios.

If the amplitude of the GWB is achieved by a larger overall SMBH mass function amplitude, then a lower radiative efficiency by around an order of magnitude is required. If the peak mass that contributes to the background is significantly lower than the value predicted by scaling relations, this also implies a larger Eddington ratio.

\begin{figure}
    \centering
    \includegraphics[width=\linewidth]{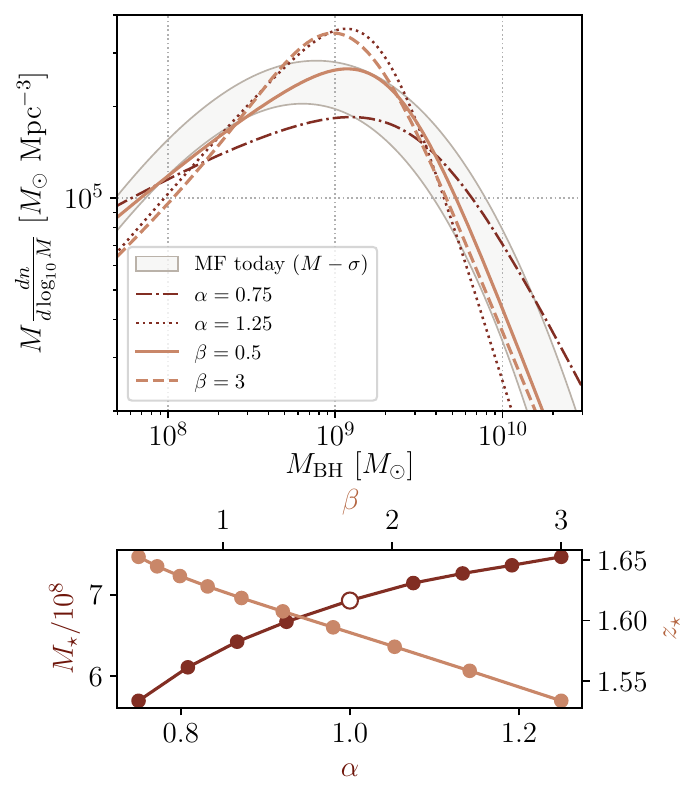}
    \caption{The top panel shows the black hole mass density per logarithmic mass, including a mass and redshift-dependent Eddington ratio. The dark red (dash-dotted and dotted lines) correspond to different power-law coefficients of the mass-dependence $\alpha$, while the orange (solid and dashed) show different values for the redshift dependence $\beta$. The bottom panel shows the pivot mass $M_{\star}$ and redshift $z_{\star}$ for different value of $\alpha$ and $\beta$ that lead to the same amplitude around $M_{\rm peak}$.}
    \label{fig:lEdd_Mz}
\end{figure}

\subsection{Mass and redshift-varying Eddington ratio}\label{sec:Ledd_M_z}
Finally, we consider that the Eddington ratio may vary with mass and redshift according to the parametrization introduced in Eq.~\ref{eq:L_Mz}, but keep the radiative efficiency fixed to the value that satisfies the total mass density inferred from the local mass function. Hence, the aforementioned parametrization of the Eddington ratio only changes how the mass density is distributed across black hole masses, but not its integral. 

The right-hand side of Eq.~\ref{eq:rhoBH_Ledd} peaks at a luminosity $L\sim 8\times 10^{12} L_{\odot}$ and therefore the peak of the mass density will be the corresponding mass for a given value of $\lambda$ (e.g., $\lambda=0.2$ corresponds to a $10^9 M_{\odot}$ peak mass density). Introducing a power-law dependence of the Eddington ratio on the mass changes the Jacobian when mapping the quasar luminosity to BH mass. From Eq.~\ref{eq:rhoBH_Ledd}, we can see that this results in a wider shape for the mass density per logarithmic mass if $\alpha<1$ and narrower if $\alpha>1$, since the same shape for the luminosity function is being mapped to a broader/narrower mass range, respectively, while the amplitude changes by a factor of $\alpha^2$ above the peak (the second term dominates). This can be seen by comparing the dark red (dashed-dotted and dotted) lines in Fig.~\ref{fig:lEdd_Mz}. 

The contribution to the mass density of a given redshift (i.e. the redshift integrand of Eq.~\ref{eq:rhoBH_Ledd}) of a $L\sim 8\times 10^{12} L_{\odot}$ quasar peaks around $z\sim 1.4$. The break luminosity $L_*$ increases between $0<z\lesssim 2$ and remains roughly constant above that while the amplitude decreases. Hence, changes to the Eddington ratio above $z\sim 2$ do not significantly change the mass function. Since the GW signal is dominated by BH masses above the peak of the mass density kernel, we may also infer that the relevant redshift (the redshift at which the integrand of Eq.~\ref{eq:rhoBH_Ledd} peaks) will necessarily within $1.4\lesssim z \lesssim 2$. Hence, so long as the pivot redshift in Eq.~\ref{eq:L_Mz} is around the redshift that most contributes to the mass density of $M^{\rm GW}_{\rm peak}$ black holes, then their mass density does not depend on $\beta$. Increasing/decreasing the Eddington ratio for $1.4<z\lesssim 2$ will decrease/increase the BH mass function above the peak. 

When comparing the local BH mass function to the accreted mass function inferred from the QLF, with or without mergers, it is useful to focus on a particular mass range, instead of the entire mass function. We focus on the mass range relevant to the gravitational wave signal, say around $M_{\rm peak}$, and require that the local and accreted mass functions match only around that mass. While the mass-dependent Eddington ratio may dramatically change the shape of the inferred mass function, it is always possible to pick a pivot mass $M_{\star}$ such that the characteristic strain at the peak $\frac{d h^2_c}{d\log_{10}M}(M_{\rm peak}) \sim M_{\rm peak}^{5/3} \frac{dn}{d\log_{10}M}$ does not depend on $\alpha$. That is, to choose $M_{\star}$ such that
\begin{equation}
    \frac{d}{d\alpha} \left.\left(\frac{d h^2_c}{d\log_{10}M}\right)\right|_{M_{\rm peak}} =0.
\end{equation}
In the results shown in Fig.~\ref{fig:lEdd_Mz}, we integrate $\frac{d h^2_c}{d\log_{10}M}$ around the $1\sigma$ width of the characteristic strain kernel (i.e. from $\log_{10} (M/M_{\odot})= 8.9-10.1$), and take the same approach was taken for the pivot redshift. The top panel of Fig.~\ref{fig:lEdd_Mz} shows a few examples of the mass function inferred from the QLF for different parameters of the mass and redshift dependence. Since $M_{\star}$ and $z_{\star}$ were picked according to the prescription above, they all agree with each other and with the local mass function around $M_{\rm peak}$. The bottom panel shows the value of the pivot mass and redshift for each value of the power law coefficients $\alpha$ and $\beta$, showing that it varies vary little as a function of each parameter.

In summary, this shows that requiring that the accreted and local mass functions agree around the peak mass that contributes to the GWB can always be expressed as a bound on the Eddington ratio for a characteristic mass $M_{\star} \sim 7\times 10^8 M_{\odot}$ and a redshift of $z_{\star} \sim 1.6$, somewhat independently of the values of $\alpha$ and $\beta$. In the presence of mergers, $M_{\star}$ is lowered by a factor of $(1 + \bar{q})^N$. Since the Eddington ratio is then increased by $(1 + \bar{q})^N$, the peak redshift will increase from 1.6 and saturate at $z=2$, as shown in the bottom panel of Fig.~\ref{fig:dpsi_dz}.

Finally, we note that introducing a scatter in the value of the Eddington ratio for a fixed luminosity will result in a larger abundance of black holes in the high-mass end. Scatter has a negligible effect in the low-mass/low-luminosity regime, where the slope of the QLF is flatter, while in the high-mass/high-luminosity end will be boosted by the quasars that scatter to a lower Eddington ratio (hence, higher mass). This will in turn require an even higher average Eddington ratio to offset this effect. Suppose that each quasar of luminosity $L$ has a probability of being hosted by a black hole of mass $M$ given by a lognormal distribution
\begin{equation}
\begin{split}
    p(\log_{10}M|\log_{10}L) =& \frac{1}{\sqrt{2\pi}\sigma_{\lambda}} \\ &\times \exp\left[-\frac{1}{2}\left(\frac{\log_{10}M - \log_{10}\bar{M}(L)}{\sigma_{\lambda}}\right)^2\right],
\end{split}
\end{equation}
where $\bar{M}(L) = L \frac{t_E} {\lambda c^2}$ and $\sigma_{\lambda}$ is the scatter of the black hole mass, and the final mass function follows exactly as in Eq.~\ref{eq:MF_scatter}, as a convolution of the mass function with no scatter with the lognormal above. Similarly to the result found in Ref.~\cite{Sato-Polito:2023gym}, the peak mass that contributes to the characteristic strain (i.e., the maximum of $M^{5/3} \psi$) scales as $\log M_{\rm peak} \propto \sigma_{\lambda}^2$. Hence, if $M_{\rm peak}$ must match the value implied by the present-day mass function and, as argued in Sec.\ref{sec:acc+mergers}, scales as $M_{\rm peak} \propto \lambda^{-1}$, we find that a larger mean Eddington ratio is required. In the absence of mergers, if $\sigma_{\lambda} = 0.3$ or $0.5$, then $\bar{\lambda} = 0.45$ or $2$.

\section{Conclusions}\label{sec:conclusion}
We explored the evolution of the supermassive black holes mass function via the continuity equation. The evolution through accretion can be connected to observations of the quasar luminosity function, the evolution through mergers to the gravitational-wave background recently detected by PTAs, and the mass function today estimated from scaling relations and galaxy catalogs provides a boundary condition.

We showed that the evolution of the most massive SMBHs (roughly $3\times 10^{9} M_{\odot}$) which are expected to produce the dominant contribution to the SGWB can be simplified into a two-step process: mergers are dominated by low redshifts $z\lesssim 1$, while accretion dominates at $1.4\lesssim z \lesssim 2$. This results in a substantial simplification of the description of the mass function evolution and enables a simple relation between mergers and the average Eddington ratio. In general, we conclude that the observation of the SGWB, provided that it is produced by the mergers of SMBHs,  implies a larger Eddington ratio for the quasar population when compared to the scenario in which mergers are negligible. 

The simplified two stage model of the evolution of the most massive SMBHs (accretion, then mergers) only depends on the assumption that SMBHs can only merge later than their host halos --- therefore the halo merger rate provides the highest redshift estimate of black hole mergers --- and that the connection between them does not change by more than roughly $\sim 3$ orders of magnitude between today and the peak of quasar activity. We neglect any time delays between halo, galaxy, and SMBH mergers, which would only lead to the merger signal being dominated by even lower redshifts. 

The conclusion that the most massive SMBHs merged after accretion was mostly complete also has  implications for theoretical predictions of their spin. While the precise number of mergers black holes with $M\gtrsim 10^9 M_{\odot}$ have undergone is highly model dependent, the values derived from the fiducial model described in Sec.~\ref{sec:merger} suggest a modest $\mathcal{O}(1)$ number.  The most massive black holes at low redshift thus likely have their spin strongly modified, and potentially dominated by, the orbital angular momentum associated with the mergers that produce the GWB.    
For example, assuming an estimate based on Ref.~\cite{2003ApJ...585L.101H} and the fiducial mass ratio distribution adopted in this work, where the average value is $\bar{q}=0.4$, the final spin is potentially non-negligible, around $\sim 0.5$, even if the initial spin is very small.   The presence of significant spin due to mergers may be important for understanding the ubiquity of jets and their associated feedback \cite{Fabian2012} in the massive black hole population at low redshift \cite{Chiaberge2011}.   

In order to explain the observed quasar population given a population of low-redshift black hole mergers, the required Eddington ratio of quasars must be larger by an amount directly related to the shift in masses by mergers; this shift can be approximated by the ratio of the final mass to the most massive progenitor $M_f/M_i = (1+\bar{q})^N \sim \lambda/0.2$, where $\lambda = 0.2$ is the best-fit Eddington ratio in the absence of mergers, $\bar{q}$ is the average mass ratio of the mergers, and $N$ is their number. This connection may pose a challenge in producing a SGWB consistent with the amplitude measured by PTAs.  Many estimates of the local BH mass function tend to underpredict the characteristic strain amplitude of the SWGB. Multiple merger events can boost the GWB amplitude, but we show that this inherently leads to a large Eddington ratio for the quasar population, in tension with measurements based on quasar linewidths (Fig. \ref{fig:Ledd_SDSS}).

Since this work focuses on the most massive black holes which dominate the contribution to the SGWB signal (of mass $M_{\rm peak}$), introducing additional mass and redshift dependence to the Eddington ratio does not significantly change the picture described above. This is due to the fact that one can always define a pivot redshift in which the relevant mass $M_{\rm peak}$ receives most of its accretion contribution, and a pivot mass such that the abundance of black holes of mass $M_{\rm peak}$ is fixed to the value given by the present-day mass function (which may be shifted by mergers). Hence, the only relevant Eddington ratio is that of quasars corresponding to the pivot mass at the pivot redshift. We also show that scatter in the relation between quasar luminosity and black hole mass leads to a higher Eddington ratio than would be inferred in the absence of scatter.

\acknowledgments
We would like to thank Phil Hopkins, Nianyi Chen, Luke Zoltan Kelley, and Nadia Zakamska for helpful discussions. GSP would like to thank Marc Kamionkowski and Lingyuan Ji for discussions several years ago about the halo/black hole coagulation equation that helped frame early stages of this work. GSP is supported by NSF PHY-2209991. MZ is supported by NSF 2209991 and NSF-BSF 2207583, and EQ is supported in part by a Simons Investigator Award from the Simons Foundation. 

\bibliography{ref.bib}

\providecommand{\href}[2]{#2}\begingroup\raggedright\begin{thebibliography}{10}

\bibitem{1998Natur.395A..14R}
D.~{Richstone}, E.~A. {Ajhar}, R.~{Bender}, G.~{Bower}, A.~{Dressler}, S.~M. {Faber}, {\em et~al.}, ``{Supermassive black holes and the evolution of galaxies.},'' \href{http://dx.doi.org/10.48550/arXiv.astro-ph/9810378}{{\em \nat} {\bfseries 385} no.~6701, (Oct., 1998) A14}, \href{http://arxiv.org/abs/astro-ph/9810378}{{\ttfamily arXiv:astro-ph/9810378 [astro-ph]}}.

\bibitem{kormendy_ho}
J.~{Kormendy} and L.~C. {Ho}, ``{Coevolution (Or Not) of Supermassive Black Holes and Host Galaxies},'' \href{http://dx.doi.org/10.1146/annurev-astro-082708-101811}{{\em \araa} {\bfseries 51} no.~1, (Aug., 2013) 511--653}, \href{http://arxiv.org/abs/1304.7762}{{\ttfamily arXiv:1304.7762 [astro-ph.CO]}}.

\bibitem{salpeter_1964}
E.~E. {Salpeter}, ``{Accretion of Interstellar Matter by Massive Objects.},'' \href{http://dx.doi.org/10.1086/147973}{{\em \apj} {\bfseries 140} (Aug., 1964) 796--800}.

\bibitem{cavaliere_1971}
A.~{Cavaliere}, P.~{Morrison}, and K.~{Wood}, ``{On Quasar Evolution},'' \href{http://dx.doi.org/10.1086/151206}{{\em \apj} {\bfseries 170} (Dec., 1971) 223}.

\bibitem{small_blandford_1992}
T.~A. {Small} and R.~D. {Blandford}, ``{Quasar evolution and the growth of black holes.},'' \href{http://dx.doi.org/10.1093/mnras/259.4.725}{{\em \mnras} {\bfseries 259} (Dec., 1992) 725--737}.

\bibitem{yu_tremaine_2002}
Q.-j. Yu and S.~Tremaine, ``{Observational constraints on growth of massive black holes},'' \href{http://dx.doi.org/10.1046/j.1365-8711.2002.05532.x}{{\em Mon. Not. Roy. Astron. Soc.} {\bfseries 335} (2002) 965--976}, \href{http://arxiv.org/abs/astro-ph/0203082}{{\ttfamily arXiv:astro-ph/0203082}}.

\bibitem{marconi_2004}
A.~Marconi, G.~Risaliti, R.~Gilli, L.~K. Hunt, R.~Maiolino, and M.~Salvati, ``{Local supermassive black holes, relics of active galactic nuclei and the x-ray background},'' \href{http://dx.doi.org/10.1111/j.1365-2966.2004.07765.x}{{\em Mon. Not. Roy. Astron. Soc.} {\bfseries 351} (2004) 169}, \href{http://arxiv.org/abs/astro-ph/0311619}{{\ttfamily arXiv:astro-ph/0311619}}.

\bibitem{merloni_2008}
A.~{Merloni} and S.~{Heinz}, ``{A synthesis model for AGN evolution: supermassive black holes growth and feedback modes},'' \href{http://dx.doi.org/10.1111/j.1365-2966.2008.13472.x}{{\em \mnras} {\bfseries 388} no.~3, (Aug., 2008) 1011--1030}, \href{http://arxiv.org/abs/0805.2499}{{\ttfamily arXiv:0805.2499 [astro-ph]}}.

\bibitem{SWM_2009}
F.~{Shankar}, D.~H. {Weinberg}, and J.~{Miralda-Escud{\'e}}, ``{Self-Consistent Models of the AGN and Black Hole Populations: Duty Cycles, Accretion Rates, and the Mean Radiative Efficiency},'' \href{http://dx.doi.org/10.1088/0004-637X/690/1/20}{{\em \apj} {\bfseries 690} no.~1, (Jan., 2009) 20--41}, \href{http://arxiv.org/abs/0710.4488}{{\ttfamily arXiv:0710.4488 [astro-ph]}}.

\bibitem{tucci_volonteri_2017}
M.~{Tucci} and M.~{Volonteri}, ``{Constraining supermassive black hole evolution through the continuity equation},'' \href{http://dx.doi.org/10.1051/0004-6361/201628419}{{\em \aap} {\bfseries 600} (Apr., 2017) A64}, \href{http://arxiv.org/abs/1603.00823}{{\ttfamily arXiv:1603.00823 [astro-ph.GA]}}.

\bibitem{SWM_2013}
F.~{Shankar}, D.~H. {Weinberg}, and J.~{Miralda-Escud{\'e}}, ``{Accretion-driven evolution of black holes: Eddington ratios, duty cycles and active galaxy fractions},'' \href{http://dx.doi.org/10.1093/mnras/sts026}{{\em \mnras} {\bfseries 428} no.~1, (Jan., 2013) 421--446}, \href{http://arxiv.org/abs/1111.3574}{{\ttfamily arXiv:1111.3574 [astro-ph.CO]}}.

\bibitem{Hirschmann:2013qfl}
M.~Hirschmann, K.~Dolag, A.~Saro, L.~Bachmann, S.~Borgani, and A.~Burkert, ``{Cosmological simulations of black hole growth: AGN luminosities and downsizing},'' \href{http://dx.doi.org/10.1093/mnras/stu1023}{{\em Mon. Not. Roy. Astron. Soc.} {\bfseries 442} no.~3, (2014) 2304--2324}, \href{http://arxiv.org/abs/1308.0333}{{\ttfamily arXiv:1308.0333 [astro-ph.CO]}}.

\bibitem{Springel:2017tpz}
V.~Springel {\em et~al.}, ``{First results from the IllustrisTNG simulations: matter and galaxy clustering},'' \href{http://dx.doi.org/10.1093/mnras/stx3304}{{\em Mon. Not. Roy. Astron. Soc.} {\bfseries 475} no.~1, (2018) 676--698}, \href{http://arxiv.org/abs/1707.03397}{{\ttfamily arXiv:1707.03397 [astro-ph.GA]}}.

\bibitem{Dave:2019yyq}
R.~Dav\'e, D.~Angl\'es-Alc\'azar, D.~Narayanan, Q.~Li, M.~H. Rafieferantsoa, and S.~Appleby, ``{Simba: Cosmological Simulations with Black Hole Growth and Feedback},'' \href{http://dx.doi.org/10.1093/mnras/stz937}{{\em Mon. Not. Roy. Astron. Soc.} {\bfseries 486} no.~2, (2019) 2827--2849}, \href{http://arxiv.org/abs/1901.10203}{{\ttfamily arXiv:1901.10203 [astro-ph.GA]}}.

\bibitem{2022MNRAS.513..670N}
Y.~{Ni}, T.~{Di Matteo}, S.~{Bird}, R.~{Croft}, Y.~{Feng}, N.~{Chen}, {\em et~al.}, ``{The ASTRID simulation: the evolution of supermassive black holes},'' \href{http://dx.doi.org/10.1093/mnras/stac351}{{\em \mnras} {\bfseries 513} no.~1, (June, 2022) 670--692}, \href{http://arxiv.org/abs/2110.14154}{{\ttfamily arXiv:2110.14154 [astro-ph.GA]}}.

\bibitem{2023MNRAS.518.2123Z}
H.~{Zhang}, P.~{Behroozi}, M.~{Volonteri}, J.~{Silk}, X.~{Fan}, P.~F. {Hopkins}, {\em et~al.}, ``{TRINITY I: self-consistently modelling the dark matter halo-galaxy-supermassive black hole connection from z = 0-10},'' \href{http://dx.doi.org/10.1093/mnras/stac2633}{{\em \mnras} {\bfseries 518} no.~2, (Jan., 2023) 2123--2163}, \href{http://arxiv.org/abs/2105.10474}{{\ttfamily arXiv:2105.10474 [astro-ph.GA]}}.

\bibitem{2000MNRAS.311..576K}
G.~{Kauffmann} and M.~{Haehnelt}, ``{A unified model for the evolution of galaxies and quasars},'' \href{http://dx.doi.org/10.1046/j.1365-8711.2000.03077.x}{{\em \mnras} {\bfseries 311} no.~3, (Jan., 2000) 576--588}, \href{http://arxiv.org/abs/astro-ph/9906493}{{\ttfamily arXiv:astro-ph/9906493 [astro-ph]}}.

\bibitem{2003ApJ...582..559V}
M.~{Volonteri}, F.~{Haardt}, and P.~{Madau}, ``{The Assembly and Merging History of Supermassive Black Holes in Hierarchical Models of Galaxy Formation},'' \href{http://dx.doi.org/10.1086/344675}{{\em \apj} {\bfseries 582} no.~2, (Jan., 2003) 559--573}, \href{http://arxiv.org/abs/astro-ph/0207276}{{\ttfamily arXiv:astro-ph/0207276 [astro-ph]}}.

\bibitem{2006MNRAS.370..645B}
R.~G. {Bower}, A.~J. {Benson}, R.~{Malbon}, J.~C. {Helly}, C.~S. {Frenk}, C.~M. {Baugh}, {\em et~al.}, ``{Breaking the hierarchy of galaxy formation},'' \href{http://dx.doi.org/10.1111/j.1365-2966.2006.10519.x}{{\em \mnras} {\bfseries 370} no.~2, (Aug., 2006) 645--655}, \href{http://arxiv.org/abs/astro-ph/0511338}{{\ttfamily arXiv:astro-ph/0511338 [astro-ph]}}.

\bibitem{2011MNRAS.411.1467K}
B.~{Kocsis} and A.~{Sesana}, ``{Gas-driven massive black hole binaries: signatures in the nHz gravitational wave background},'' \href{http://dx.doi.org/10.1111/j.1365-2966.2010.17782.x}{{\em \mnras} {\bfseries 411} no.~3, (Mar., 2011) 1467--1479}, \href{http://arxiv.org/abs/1002.0584}{{\ttfamily arXiv:1002.0584 [astro-ph.CO]}}.

\bibitem{2012MNRAS.426..237H}
M.~{Hirschmann}, R.~S. {Somerville}, T.~{Naab}, and A.~{Burkert}, ``{Origin of the antihierarchical growth of black holes},'' \href{http://dx.doi.org/10.1111/j.1365-2966.2012.21626.x}{{\em \mnras} {\bfseries 426} no.~1, (Oct., 2012) 237--257}, \href{http://arxiv.org/abs/1206.6112}{{\ttfamily arXiv:1206.6112 [astro-ph.CO]}}.

\bibitem{NANOGrav_stoc}
{\bfseries NANOGrav} Collaboration, G.~Agazie {\em et~al.}, ``{The NANOGrav 15 yr Data Set: Evidence for a Gravitational-wave Background},'' \href{http://dx.doi.org/10.3847/2041-8213/acdac6}{{\em Astrophys. J. Lett.} {\bfseries 951} no.~1, (2023) L8}, \href{http://arxiv.org/abs/2306.16213}{{\ttfamily arXiv:2306.16213 [astro-ph.HE]}}.

\bibitem{EPTA_stoc}
{\bfseries EPTA} Collaboration, J.~Antoniadis {\em et~al.}, ``{The second data release from the European Pulsar Timing Array III. Search for gravitational wave signals},'' \href{http://arxiv.org/abs/2306.16214}{{\ttfamily arXiv:2306.16214 [astro-ph.HE]}}.

\bibitem{PPTA_stoc}
D.~J. Reardon {\em et~al.}, ``{Search for an Isotropic Gravitational-wave Background with the Parkes Pulsar Timing Array},'' \href{http://dx.doi.org/10.3847/2041-8213/acdd02}{{\em Astrophys. J. Lett.} {\bfseries 951} no.~1, (2023) L6}, \href{http://arxiv.org/abs/2306.16215}{{\ttfamily arXiv:2306.16215 [astro-ph.HE]}}.

\bibitem{CPTA_stoc}
H.~Xu {\em et~al.}, ``{Searching for the Nano-Hertz Stochastic Gravitational Wave Background with the Chinese Pulsar Timing Array Data Release I},'' \href{http://dx.doi.org/10.1088/1674-4527/acdfa5}{{\em Res. Astron. Astrophys.} {\bfseries 23} no.~7, (2023) 075024}, \href{http://arxiv.org/abs/2306.16216}{{\ttfamily arXiv:2306.16216 [astro-ph.HE]}}.

\bibitem{Hopkins_2007}
P.~F. Hopkins, G.~T. Richards, and L.~Hernquist, ``{An Observational Determination of the Bolometric Quasar Luminosity Function},'' \href{http://dx.doi.org/10.1086/509629}{{\em Astrophys. J.} {\bfseries 654} (2007) 731--753}, \href{http://arxiv.org/abs/astro-ph/0605678}{{\ttfamily arXiv:astro-ph/0605678}}.

\bibitem{shen_2020}
X.~Shen, P.~F. Hopkins, C.-A. Faucher-Gigu\`ere, D.~M. Alexander, G.~T. Richards, N.~P. Ross, and R.~C. Hickox, ``{The bolometric quasar luminosity function at z = 0\textendash{}7},'' \href{http://dx.doi.org/10.1093/mnras/staa1381}{{\em Mon. Not. Roy. Astron. Soc.} {\bfseries 495} no.~3, (2020) 3252--3275}, \href{http://arxiv.org/abs/2001.02696}{{\ttfamily arXiv:2001.02696 [astro-ph.GA]}}.

\bibitem{Sato-Polito:2023gym}
G.~Sato-Polito, M.~Zaldarriaga, and E.~Quataert, ``{Where are the supermassive black holes measured by PTAs?},'' \href{http://dx.doi.org/10.1103/PhysRevD.110.063020}{{\em Phys. Rev. D} {\bfseries 110} no.~6, (2024) 063020}, \href{http://arxiv.org/abs/2312.06756}{{\ttfamily arXiv:2312.06756 [astro-ph.CO]}}.

\bibitem{Sato-Polito:2024lew}
G.~Sato-Polito and M.~Zaldarriaga, ``{The distribution of the gravitational-wave background from supermassive black holes},'' \href{http://arxiv.org/abs/2406.17010}{{\ttfamily arXiv:2406.17010 [astro-ph.CO]}}.

\bibitem{soltan1982}
A.~{Soltan}, ``{Masses of quasars.},'' \href{http://dx.doi.org/10.1093/mnras/200.1.115}{{\em \mnras} {\bfseries 200} (July, 1982) 115--122}.

\bibitem{mcconnell_ma}
N.~J. {McConnell} and C.-P. {Ma}, ``{Revisiting the Scaling Relations of Black Hole Masses and Host Galaxy Properties},'' \href{http://dx.doi.org/10.1088/0004-637X/764/2/184}{{\em \apj} {\bfseries 764} no.~2, (Feb., 2013) 184}, \href{http://arxiv.org/abs/1211.2816}{{\ttfamily arXiv:1211.2816 [astro-ph.CO]}}.

\bibitem{bernardi_VDF}
M.~{Bernardi}, F.~{Shankar}, J.~B. {Hyde}, S.~{Mei}, F.~{Marulli}, and R.~K. {Sheth}, ``{Galaxy luminosities, stellar masses, sizes, velocity dispersions as a function of morphological type},'' \href{http://dx.doi.org/10.1111/j.1365-2966.2010.16425.x}{{\em \mnras} {\bfseries 404} no.~4, (June, 2010) 2087--2122}, \href{http://arxiv.org/abs/0910.1093}{{\ttfamily arXiv:0910.1093 [astro-ph.CO]}}.

\bibitem{shankar_2009}
F.~{Shankar}, D.~H. {Weinberg}, and J.~{Miralda-Escud{\'e}}, ``{Self-Consistent Models of the AGN and Black Hole Populations: Duty Cycles, Accretion Rates, and the Mean Radiative Efficiency},'' \href{http://dx.doi.org/10.1088/0004-637X/690/1/20}{{\em \apj} {\bfseries 690} no.~1, (Jan., 2009) 20--41}, \href{http://arxiv.org/abs/0710.4488}{{\ttfamily arXiv:0710.4488 [astro-ph]}}.

\bibitem{liepold_ma_2024}
E.~R. Liepold and C.-P. Ma, ``{Big Galaxies and Big Black Holes: The Massive Ends of the Local Stellar and Black Hole Mass Functions and the Implications for Nanohertz Gravitational Waves},'' \href{http://dx.doi.org/10.3847/2041-8213/ad66b8}{{\em Astrophys. J. Lett.} {\bfseries 971} no.~2, (2024) L29}, \href{http://arxiv.org/abs/2407.14595}{{\ttfamily arXiv:2407.14595 [astro-ph.GA]}}.

\bibitem{benson_2004}
A.~J. Benson, M.~Kamionkowski, and S.~H. Hassani, ``{Self-consistent theory of halo mergers},'' \href{http://dx.doi.org/10.1111/j.1365-2966.2005.08679.x}{{\em Mon. Not. Roy. Astron. Soc.} {\bfseries 357} (2005) 847--858}, \href{http://arxiv.org/abs/astro-ph/0407136}{{\ttfamily arXiv:astro-ph/0407136}}.

\bibitem{erickcek_2006}
A.~L. Erickcek, M.~Kamionkowski, and A.~J. Benson, ``{Supermassive Black Hole Merger Rates: Uncertainties from Halo Merger Theory},'' \href{http://dx.doi.org/10.1111/j.1365-2966.2006.10838.x}{{\em Mon. Not. Roy. Astron. Soc.} {\bfseries 371} (2006) 1992--2000}, \href{http://arxiv.org/abs/astro-ph/0604281}{{\ttfamily arXiv:astro-ph/0604281}}.

\bibitem{Ellis:2023dgf}
J.~Ellis, M.~Fairbairn, G.~H\"utsi, J.~Raidal, J.~Urrutia, V.~Vaskonen, and H.~Veerm\"ae, ``{Gravitational waves from supermassive black hole binaries in light of the NANOGrav 15-year data},'' \href{http://dx.doi.org/10.1103/PhysRevD.109.L021302}{{\em Phys. Rev. D} {\bfseries 109} no.~2, (2024) L021302}, \href{http://arxiv.org/abs/2306.17021}{{\ttfamily arXiv:2306.17021 [astro-ph.CO]}}.

\bibitem{Ellis:2023owy}
J.~Ellis, M.~Fairbairn, G.~H\"utsi, M.~Raidal, J.~Urrutia, V.~Vaskonen, and H.~Veerm\"ae, ``{Prospects for future binary black hole gravitational wave studies in light of PTA measurements},'' \href{http://dx.doi.org/10.1051/0004-6361/202346268}{{\em Astron. Astrophys.} {\bfseries 676} (2023) A38}, \href{http://arxiv.org/abs/2301.13854}{{\ttfamily arXiv:2301.13854 [astro-ph.CO]}}.

\bibitem{2020A&A...634A.135G}
G.~{Girelli}, L.~{Pozzetti}, M.~{Bolzonella}, C.~{Giocoli}, F.~{Marulli}, and M.~{Baldi}, ``{The stellar-to-halo mass relation over the past 12 Gyr. I. Standard {\ensuremath{\Lambda}}CDM model},'' \href{http://dx.doi.org/10.1051/0004-6361/201936329}{{\em \aap} {\bfseries 634} (Feb., 2020) A135}, \href{http://arxiv.org/abs/2001.02230}{{\ttfamily arXiv:2001.02230 [astro-ph.CO]}}.

\bibitem{Chen:2018znx}
S.~Chen, A.~Sesana, and C.~J. Conselice, ``{Constraining astrophysical observables of Galaxy and Supermassive Black Hole Binary Mergers using Pulsar Timing Arrays},'' \href{http://dx.doi.org/10.1093/mnras/stz1722}{{\em Mon. Not. Roy. Astron. Soc.} {\bfseries 488} no.~1, (2019) 401--418}, \href{http://arxiv.org/abs/1810.04184}{{\ttfamily arXiv:1810.04184 [astro-ph.GA]}}.

\bibitem{2014MNRAS.443..874B}
M.~{Bernardi}, A.~{Meert}, V.~{Vikram}, M.~{Huertas-Company}, S.~{Mei}, F.~{Shankar}, and R.~K. {Sheth}, ``{Systematic effects on the size-luminosity relations of early- and late-type galaxies: dependence on model fitting and morphology},'' \href{http://dx.doi.org/10.1093/mnras/stu1106}{{\em \mnras} {\bfseries 443} no.~1, (Sept., 2014) 874--897}.

\bibitem{Sesana:2016yky}
A.~Sesana, F.~Shankar, M.~Bernardi, and R.~K. Sheth, ``{Selection bias in dynamically measured supermassive black hole samples: consequences for pulsar timing arrays},'' \href{http://dx.doi.org/10.1093/mnrasl/slw139}{{\em Mon. Not. Roy. Astron. Soc.} {\bfseries 463} no.~1, (2016) L6--L11}, \href{http://arxiv.org/abs/1603.09348}{{\ttfamily arXiv:1603.09348 [astro-ph.GA]}}.

\bibitem{QSO_SDSS}
Q.~Wu and Y.~Shen, ``{A Catalog of Quasar Properties from Sloan Digital Sky Survey Data Release 16},'' \href{http://dx.doi.org/10.3847/1538-4365/ac9ead}{{\em Astrophys. J. Suppl.} {\bfseries 263} no.~2, (2022) 42}, \href{http://arxiv.org/abs/2209.03987}{{\ttfamily arXiv:2209.03987 [astro-ph.GA]}}.

\bibitem{Liepold:2024woa}
E.~R. Liepold and C.-P. Ma, ``{Big Galaxies and Big Black Holes: The Massive Ends of the Local Stellar and Black Hole Mass Functions and the Implications for Nanohertz Gravitational Waves},'' \href{http://dx.doi.org/10.3847/2041-8213/ad66b8}{{\em Astrophys. J. Lett.} {\bfseries 971} no.~2, (2024) L29}, \href{http://arxiv.org/abs/2407.14595}{{\ttfamily arXiv:2407.14595 [astro-ph.GA]}}.

\bibitem{2013BASI...41...61S}
Y.~{Shen}, ``{The mass of quasars},'' \href{http://dx.doi.org/10.48550/arXiv.1302.2643}{{\em Bulletin of the Astronomical Society of India} {\bfseries 41} no.~1, (Mar., 2013) 61--115}, \href{http://arxiv.org/abs/1302.2643}{{\ttfamily arXiv:1302.2643 [astro-ph.CO]}}.

\bibitem{DS15}
R.~{D'Souza}, S.~{Vegetti}, and G.~{Kauffmann}, ``{The massive end of the stellar mass function},'' \href{http://dx.doi.org/10.1093/mnras/stv2234}{{\em \mnras} {\bfseries 454} no.~4, (Dec., 2015) 4027--4036}, \href{http://arxiv.org/abs/1509.07418}{{\ttfamily arXiv:1509.07418 [astro-ph.GA]}}.

\bibitem{2003ApJ...585L.101H}
S.~A. {Hughes} and R.~D. {Blandford}, ``{Black Hole Mass and Spin Coevolution by Mergers},'' \href{http://dx.doi.org/10.1086/375495}{{\em \apjl} {\bfseries 585} no.~2, (Mar., 2003) L101--L104}, \href{http://arxiv.org/abs/astro-ph/0208484}{{\ttfamily arXiv:astro-ph/0208484 [astro-ph]}}.

\bibitem{Fabian2012}
A.~C. {Fabian}, ``{Observational Evidence of Active Galactic Nuclei Feedback},'' \href{http://dx.doi.org/10.1146/annurev-astro-081811-125521}{{\em \araa} {\bfseries 50} (Sept., 2012) 455--489}, \href{http://arxiv.org/abs/1204.4114}{{\ttfamily arXiv:1204.4114 [astro-ph.CO]}}.

\bibitem{Chiaberge2011}
M.~{Chiaberge} and A.~{Marconi}, ``{On the origin of radio loudness in active galactic nuclei and its relationship with the properties of the central supermassive black hole},'' \href{http://dx.doi.org/10.1111/j.1365-2966.2011.19079.x}{{\em \mnras} {\bfseries 416} no.~2, (Sept., 2011) 917--926}, \href{http://arxiv.org/abs/1105.4889}{{\ttfamily arXiv:1105.4889 [astro-ph.CO]}}.

\bibitem{PS_1974}
W.~H. {Press} and P.~{Schechter}, ``{Formation of Galaxies and Clusters of Galaxies by Self-Similar Gravitational Condensation},'' \href{http://dx.doi.org/10.1086/152650}{{\em \apj} {\bfseries 187} (Feb., 1974) 425--438}.

\bibitem{BCEK_1991}
J.~R. {Bond}, S.~{Cole}, G.~{Efstathiou}, and N.~{Kaiser}, ``{Excursion Set Mass Functions for Hierarchical Gaussian Fluctuations},'' \href{http://dx.doi.org/10.1086/170520}{{\em \apj} {\bfseries 379} (Oct., 1991) 440}.

\bibitem{lacey_cole_1993}
C.~{Lacey} and S.~{Cole}, ``{Merger rates in hierarchical models of galaxy formation},'' \href{http://dx.doi.org/10.1093/mnras/262.3.627}{{\em \mnras} {\bfseries 262} no.~3, (June, 1993) 627--649}.

\end{thebibliography}\endgroup
\bibliographystyle{utcaps}

\appendix

\section{EPS mergers}\label{app:EPS}
In the spherical collapse model, a region with a density field linearly extrapolated to the present time $\delta_0(\pmb{x})$ will have collapsed to form a virialized object at a time $t$ if $\delta_0(\pmb{x}) >\delta_c/D(t) \equiv \delta_c(t)$, where $D(t)$ is the linear growth rate normalized to unity today. In order to assign a mass to the collapsed regions, we consider the density field smoothed over a spherically symmetric window function $W$:
\begin{equation}
    \tilde{\delta}(\pmb{x}; R) = \int d^3 \pmb{x}' \delta_0(\pmb{x}') W(|\pmb{x}-\pmb{x}'|; R),
\end{equation}
which has an associated mass of $M\sim \rho_0 R^3$. The Press-Schechter formalism \cite{PS_1974} equates the probability of $\tilde{\delta}(\pmb{x}; R)$ surpassing the barrier $\delta_c(t)$ with the fraction of collapsed objects above the corresponding mass. The resulting halo mass function is given by
\begin{equation}
    \frac{dn}{d\ln M} (M,z) = \sqrt{\frac{2}{\pi}} \frac{\rho_0}{M} \left|\frac{d\ln \sigma}{d\ln M} \right|\frac{\delta_c(z)}{\sigma(M)} e^{-\frac{\delta_c(z)^2}{2\sigma^2(M)}},
    \label{eq:HMF}
\end{equation}
where $\sigma$ is the variance of the smoothed density field.

The extended Press-Schechter (EPS) \cite{BCEK_1991, lacey_cole_1993} offers an alternative derivation of the growth of structure. Since $\sigma^2(M)$ is a monotonically decreasing function of $M$, it can be used as the mass variable. Each location $\pmb{x}$ then corresponds to a trajectory of $\tilde{\delta}$ as a function of $\sigma^2$, the value of the density field at that location when smoother over a filter of mass $M$ (corresponding to $\sigma^2)$. This allows a self-consistent prediction of the mass function as well as the merger rate of dark matter halos. 

The probability per unit time of a halo of mass $M_2$ will merge with a halo of mass $M_1=M-M_2$ to form a final halo of mass $M$ is given by
\begin{equation}
\begin{split}
    \frac{dp}{dM_2 dt} =& \sqrt{\frac{2}{\pi}} \frac{1}{M} \left| \frac{\dot{\delta}_c}{\delta_c}\right| \left| \frac{d\ln \sigma}{d\ln M}\right| \left[ 1 - \frac{\sigma^2(M)}{\sigma^2(M_1)}\right]^{-3/2} \frac{\delta_c(t)}{\sigma(M)} \times \\ &\exp\left\{ -\frac{\delta_c^2(t)}{2} \left( \frac{1}{\sigma^2(M)} - \frac{1}{\sigma^2(M_1)} \right) \right\}.
    \label{eq:dpdmdt}
\end{split}
\end{equation}
The total number of mergers per unit time per unit volume between halos of mass $M_2$ and $M_1$ is therefore
\begin{equation}
    \frac{d^3n_h}{dM_1 dM_2 dt} = \frac{dp}{dM_2 dt} \frac{dn}{dM_1},
    \label{eq:merger_rate}
\end{equation}

The number of mergers per final halo mass can be computed from Eqs.~\ref{eq:HMF}, \ref{eq:dpdmdt}, and \ref{eq:merger_rate}:
\begin{equation}
\begin{split}
    \left(\frac{dn}{dM_h}\right)^{-1} \frac{d^3 n}{dM_t dq dz} = \sqrt{\frac{2}{\pi}} \frac{M_t^2}{(1+q)^2} \frac{1}{M_1^2 \sigma(M_1)} \times & \\ \left| \frac{d\log \sigma}{d\log M_h} (M_t)\right| \left| \frac{d\delta_c}{dz}\right| \left[ 1- \frac{\sigma^2(M_t)}{\sigma^2(M_1)}\right]^{-3/2}&
\end{split}
\end{equation}
where $M_1=M_t/(1+q)$. For a power spectrum $P(k) \propto k^n$, we find that the mass variance scales as $\sigma^2(M) \propto M^{-(3+n)/3}$ and the number of mergers per halo thus scales as
\begin{equation}
\begin{split}
\left(\frac{dn}{dM_h}\right)^{-1} \frac{d^3 n}{dM_t dq dz} \propto M_t^{\alpha} \frac{d\delta_c}{dz} (1+q)^{\alpha} \left[1- (1+q)^{2\alpha}\right]
\end{split}
\end{equation}
where $\alpha=-(3+n)/6$. When $n\rightarrow -3$, the number of mergers per halo is indeed universal, and only acquires a small mass dependence for any relevant value of $n$. We also note that the number of mergers per halo increases with redshifts due to the dependence on $d\delta_c/dz$. 

\end{document}